\documentclass[prl,amssymb,twocolumn,showpacs,superscriptaddress]{revtex4}
\usepackage{graphicx}
\usepackage{bm}

\begin{document}

\title{Effective temperature in driven vortex lattices with 
random pinning}

\author{Alejandro B. Kolton}
\affiliation{Centro At{\'{o}}mico Bariloche, 8400 S. C. de Bariloche,
R{\'{\i}}o Negro, Argentina.}
\author{Raphael Exartier}
\affiliation{LMDH, 4. Place Jussieu, 75252 Paris Cedex 05, France.}
\author{Leticia F. Cugliandolo}
\affiliation{LPT Ecole Normale Sup{\'e}rieure , 24 rue Lhomond, 75231
Paris Cedex 05 France and \\
LPTHE, 4 Place Jussieu,  75252 Paris Cedex 05, France.}
\author{Daniel Dom{\'{\i}}nguez}
\affiliation{Centro At{\'{o}}mico Bariloche, 8400 S. C. de Bariloche,
R{\'{\i}}o Negro, Argentina.}
\author{N. Gr{\o}nbech-Jensen}
\affiliation{Department of Applied Science, University of California, 
Davis, California 95616 and \\ NERSC, Lawrence Berkeley National Laboratory, 
Berkeley, California 94720, USA}

\begin{abstract}
We study numerically correlation and response functions 
in non-equilibrium driven
vortex lattices with random pinning.
From a generalized  fluctuation-dissipation relation 
we calculate an effective
transverse temperature  in the fluid moving phase. 
We find that the effective temperature 
decreases with increasing driving force and
becomes equal to the equilibrium melting temperature  
when the dynamic transverse freezing occurs.
We also discuss how the effective temperature can be measured experimentally
from a generalized Kubo formula.
\end{abstract}


\maketitle

Whether and how can one extend 
thermodynamic concepts to nonequilibrium
systems is a very important challenge in theoretical physics.
Many definitions of nonequilibrium temperatures 
have been proposed in different contexts but it has been rarely 
checked if they conform with 
the expected properties of a temperature.

Cugliandolo, Kurchan and Peliti~\cite{CKP} have 
introduced the notion of time-scale dependent 
``effective temperatures'' $T_{\tt eff}$ from a
modification of the fluctuation-dissipation theorem (FDT)
in slowly evolving out of equilibrium systems.
$T_{\tt eff}$  is defined 
from the slope of the parametric plot of the 
integrated response against the correlation function
of a given pair of observables when the latter is bounded or,
equivalently, it is the inverse of 
twice the slope of the parametric plot of the 
integrated response against the displacement
when the correlation is unbounded.
This definition yields a bona fide temperature in the thermodynamic sense 
since it can be measured with a thermometer, it controls the direction 
of heat flow for a given time scale and it satisfies a
zero-th law within each time scale.
$T_{\tt eff}$ was 
found analytically in mean-field glassy models~\cite{CK}
and it was  successfully studied in structural and spin
glasses, both numerically~\cite{teff_sim}
and experimentally \cite{teff_exp}, in granular matter
\cite{teff_gran,teff_gran2} and in weakly driven sheared fluids
\cite{driven_teff,shear}.

In their study  of  driven vortex lattices in type II superconductors, 
Koshelev and Vinokur \cite{KV} have defined a
``shaking'' temperature $T_{\tt sh}$ from the fluctuating force 
felt by a vortex
configuration moving in a random pinning potential landscape.
This lead to the prediction of a dynamic phase transition
between a liquid-like phase of vortices moving at low driving forces 
and a crystalline vortex lattice moving at
large forces, when $T_{\tt sh}$  equals the
equilibrium melting temperature of the vortex system \cite{KV,AKV}.
However, later work \cite{GLD,BMR,SV} has shown that the perturbation 
theory used in \cite{KV} breaks down and that the vortex phase at large 
velocities can be an anisotropic transverse glass instead of a crystal.
In spite of this, the  shaking temperature introduced in \cite{KV}
has been a useful qualitative concept, at least phenomenologically. 
Indeed, the dynamic transitions and moving vortex phases
discussed in \cite{KV,AKV,GLD,BMR,SV} have been observed 
experimentally \cite{EXP} and in numerical simulations \cite{SIM,kolton}.

In this Letter we apply the definition of $T_{\tt eff}$ based on the 
modifications of the FDT to driven vortex lattices
above the critical force
and within the fluid moving phase. We compare our results with 
the shaking temperature of \cite{KV} and discuss  how to obtain $T_{\tt eff}$
experimentally from measurements of transverse voltage
noise and transverse resistance~\cite{Yoshino}.

The equation of motion of a vortex in position ${\bf R}_i$ is:
\begin{eqnarray*}
\eta \frac{d{\bf R}_i}{dt} = -\sum_{j\not= i}{\bm\nabla}_i U_v(R_{ij})
-\sum_p{\bm \nabla}_i U_p(R_{ip}) + {\bf F} + 
{\bm \zeta}_i(t),
\end{eqnarray*} 
where $R_{ij}=|{\bf R}_i-{\bf R}_j|$ is the distance between vortices $i,j$,
$R_{ip}=|{\bf R}_i-{\bf R}_p|$ is the distance between the vortex $i$ and
a pinning site at ${\bf R}_p$, $\eta$ is the
Bardeen-Stephen friction, 
and ${\bf F}= \frac{d\Phi_0}{c}{\bf J}\times   {\bf z}$ 
is the driving force due to a uniform 
current density ${\bf J}$. The effect of a thermal bath 
at temperature $T$ is given by the stochastic
force ${\bm\zeta}_i(t)$,  satisfying $\langle {\zeta}^\mu _i(t) 
\rangle=0$ and $\langle {\zeta}^\mu   _i(t){\zeta}^{\mu '}_j(t') 
\rangle = 2 \eta k_B T \delta(t-t') \delta_{ij} \delta_{\mu \mu '}$, 
where  $\langle \ldots \rangle$ denotes average over the ensemble 
of ${\bm\zeta}_i$. We model a 2D thin film superconductor of thickness
$d$ and size $L$ by considering a logarithmic vortex-vortex interaction 
potential:
$U_v(r)=-A_v\ln(r/\Lambda)$, with $A_v=\Phi_0^2/8\pi\Lambda$
and $\Lambda=2\lambda^2/d > L$ \cite{kolton}.
The vortices interact with a random distribution of
attractive pinning centers with 
$U_p(r)=-A_p e^{-(r/r_p)^2}$. 
Length is normalized by $r_p$, energy by $A_v$, 
and time by  $\tau=\eta r_p^2/A_v$.  
We consider $N_v$ vortices and $N_p$ pinning
centers in a rectangular box of size $L_x\times   L_y$. 
Moving vortices induce an electric field  ${\bf
E}=\frac{B}{c}{\bf V}\times {\bf z}$, with ${\bf V}=\frac{1}{N_v}\sum_i 
d{\bf R}_i/dt$. 

To study the fluctuation-dissipation relation (FDR), 
we proceed in a similar way as in previous simulations of 
structural glasses \cite{teff_sim} and consider the observables:
\begin{equation}
A_\mu   (t)=\frac{1}{N_v}\sum_{i=1}^{N_v} s_i  r^\mu   _i(t)\;;\;\;\;\;
B_\mu   (t)=\sum_{i=1}^{N_v} s_i  r^\mu   _i(t) \; ,
\end{equation}
where $s_i=-1, 1$ are random numbers with 
$\overline{s_i}=0$ and $\overline{s_i s_j}=\delta_{ij}$, 
and $r^\mu   _i=R^\mu   _i-R^\mu   _{cm}$
with $\mu   =x,y$ and
${\bf R}_{cm}$  being the center of mass coordinate. 
Taking ${\bf F}=F{\bf y}$ 
we  study separately the FDR in the transverse and 
parallel directions with respect to ${\bf F}$. 
The time correlation function between the observables 
$A_\mu   $ and $B_\mu   $ is 
\begin{equation}
C_\mu   (t,t_0) = \overline{\langle A_\mu   (t)B_\mu   (t_0)\rangle} 
=  \frac{1}{N_v}
\sum_{i=1}^{N_v}\langle 
r^\mu   _i(t) r^\mu   _i(t_0)\rangle \; ,
\end{equation}
since the $r^\mu _i$ are independent of the $s_i$. The integrated response 
function $\chi_\mu   $ for the observable $A_\mu   $ is obtained by applying 
a perturbative force ${\bf f}^\mu   _i = \epsilon s_i \hat{\bm\mu   }$ 
(where $\hat{\bm\mu   }=\hat{\bm x}, \hat{\bm y}$)
at time $t_0$ and keeping it constant for all subsequent times
on each vortex:
\begin{eqnarray}
\chi_\mu   (t,t_0) = \lim_{\epsilon \to 0}\frac{1}{\epsilon}
\Bigl[ \overline{\langle {A_\mu   (t)} \rangle}_{\epsilon} - 
\overline{\langle {A_\mu   (t)} \rangle}_{\epsilon=0}
\Bigr] \; .
\end{eqnarray}
We then define a function, $T^\mu   _{\tt eff}(t,t_0)$, by the relation,  
\begin{eqnarray}
\chi_\mu   (t,t_0)=\frac{1}{2k_B T^\mu   _{\tt eff}(t,t_0)} \Delta_\mu 
(t,t_0) \; ,
\end{eqnarray}
where
$
\Delta_\mu   (t,t_0) = \frac{1}{N_v}\sum_{i=1}^{N_v}
\langle|r^\mu   _i(t)-r^\mu   _i(t_0)|^2\rangle 
= C_\mu   (t,t)+C_\mu   (t_0,t_0)-2C_\mu   (t,t_0)
$
is the quadratic mean displacement 
in the direction of $\hat{\bm\mu   }$. 
For a system in equilibrium at temperature $T$ the FDT
requires that $T^x_{\tt eff}=T^y_{\tt eff}=T$. 
In a nonequilibrium system, like the driven 
vortex lattice with pinning, the FDT does not apply. 
Since we are interested in the {\it stationary} states reached 
by the driven vortex lattice, where aging effects are stopped
\cite{driven_teff,shear,Pierre},
then all observables  depend only on the difference
$t-t_0$, if  
we choose $t_0$ long enough to ensure stationarity. 
From the parametric plot of $\chi_\mu   (t)$ vs.
$\Delta_\mu   (t)$ one defines the effective temperature 
$T^\mu   _{\tt eff}(t)$ using Eq.~(4),
provided $T^\mu   _{\tt eff}(t)$ is a constant
in each time-scale  \cite{CKP}.

We study the transverse and longitudinal FDR for the 
moving vortex lattice as a function of driving force, $F$, 
solving the dynamic equations 
for different values of $A_p$, $n_v$, and $T$. The 
simulations are performed with pinning density 
$n_p=N_pr_p^2/L_xL_y=0.14$ in a 
box with $L_x/L_y=\sqrt{3}/2$ and $N_v=256$.
We consider $A_p/A_v=0.35, 0.2, 0.25, 0.1$, 
$n_v=N_vr_p^2/L_xL_y=0.05, 0.07$, 
and $T\leq 0.01$. We impose periodic boundary 
conditions with the algorithm of 
Ref.~\cite{log}. 
Averages are evaluated during 80000 steps of $\Delta t=0.1 \tau$
after 65536 steps for reaching stationarity. 
To calculate the response function 
$\chi_\mu   (t)$, given by Eq.~(3), we simulate two replicas of the system, 
with the perturbative force ${\bf f}^\mu   _i = \epsilon s_i \hat{\bm\mu   }$ 
applied to one of them. 
Starting from the same initial condition we let the perturbed and unperturbed 
system evolve for $5000$ time steps and calculate
${A_\mu   (t)}_{\epsilon}$ 
and ${A_\mu   (t)}_{\epsilon=0}$ respectively. 
The replicas then evolve 
again after changing the realization of the random factors, $s_i$, 
and taking the final configuration of the unperturbed system as the 
new common initial condition. From this  we get 
$\overline{\langle {A_\mu   (t)} \rangle}_{\epsilon}$, 
both for $\epsilon = 0$ and $\epsilon \neq 0$, and thereby 
the response function, $\chi_\mu   (t)$, is determined. We choose 
$\epsilon$ small enough in each case to assure a linear response of 
$\langle {A_\mu   (t)} \rangle_\epsilon$.    

At $T=0$, there are three different dynamical regimes \cite{kolton} 
when increasing $F$ 
above the critical depinning force $F_c$: two fluid phases
with plastic flow for $F_c < F <F_p$ and smectic flow 
for $F_p < F <F_t$, and a transverse solid for $F > F_t$. 
For fixed pinning density $n_p$, the characteristic 
forces $F_c$, $F_p$ and $F_t$ depend on the disorder strength $A_p$ and 
vortex density $n_v$. We start by analyzing the FDR  
in the smectic flow regime for different values of $A_p$, $n_v$ and $T$. 
\begin{figure}
\centerline{
\includegraphics[width=8.5cm]{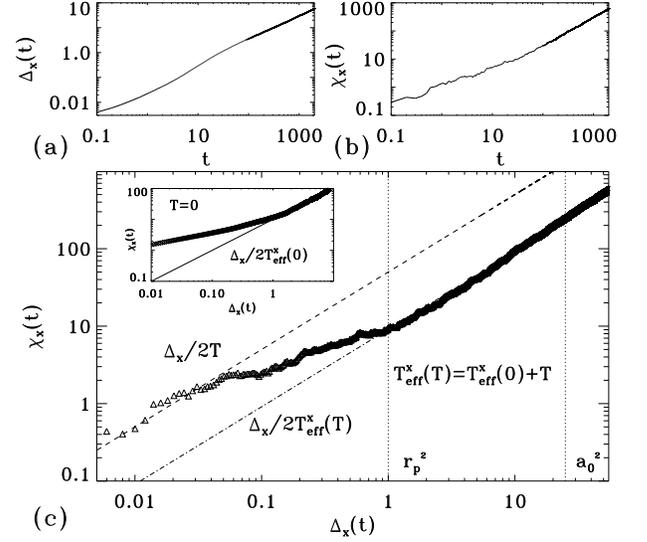}}
\caption{Averaged transverse quadratic mean displacement $\Delta_x(t)$ (a), 
and transverse response function $\chi_x(t)$ (b), for $A_p=0.2$, $T=0.01$ 
and $V=0.11$, in the smectic flow regime. (c) Transverse fluctuation 
dissipation relation $\chi_x(t)$ vs $\Delta_x(t)$ for $T=0.01$. 
The dashed line indicates $\Delta_x/2T$, with $T=0.01$ the thermal 
bath temperature. $T_{\tt eff}(T)$ is obtained from the linear 
fit to $\chi_x(t)$ for $\Delta_x>r_p^2$ (dashed dotted). We find that 
$T_{\tt eff}^x(T) \approx T_{\tt eff}^x(0)+T$, where $T_{\tt eff}^x(0)$ is the 
corresponding transverse effective temperature for $T=0$ (inset).}  
\end{figure}\noindent
In Fig.~1(a) we show the typical transverse 
quadratic mean displacements, $\Delta_x(t)$, and in Fig.~1(b) the
integrated transverse response, 
$\chi_x(t)$, for this dynamical regime. In Fig.~1(c) we show the 
FDR parametric plot of $\chi_x(t)$ against
$\Delta_x(t)$. We see that the equilibrium FDT does not apply in general
but two approximate 
linear relations exist for $\Delta_x(t)< 0.05 r_p^2$ and 
for $ \Delta_x(t) > r_p^2$, with a non-linear crossover between them. 
Following Eq.~(4), we 
find that the short displacements region corresponds to the bath temperature 
$T=0.01$, and therefore the equilibrium FDT applies in the transverse direction 
only for short times,  $t\ll   r_p/v$. 
For the large displacements region we get an effective transverse 
temperature $T_{\tt eff}^x(T)=0.045 > T$. 
In the inset of Fig.~1(a) we show the FDR for $T=0$. 
Comparing the results for different $T$, we find  
$T_{\tt eff}^x(T) \approx T_{\tt eff}^x(0)+T$. 
\begin{figure}
\centerline{
\includegraphics[width=8.5cm]{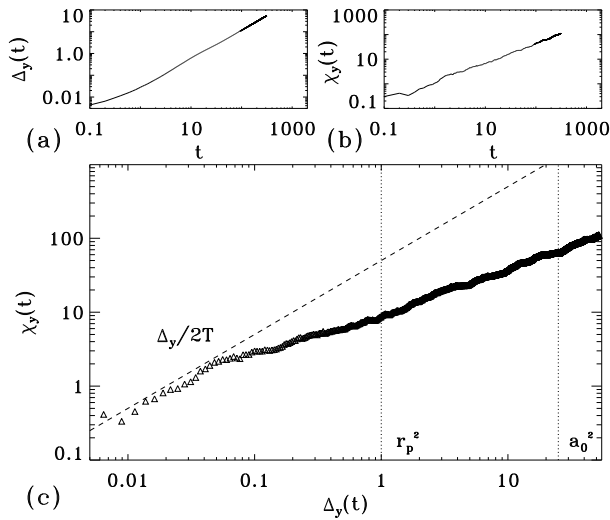}}
\caption{Averaged longitudinal quadratic mean displacement $\Delta_y(t)$ (a), 
and longitudinal response function $\chi_y(t)$ (b), for $A_p=0.2$, $T=0.01$ 
and $V=0.11$, in the smectic flow regime. (c) Longitudinal fluctuation 
dissipation relation $R_y(t)$ vs $\Delta_y(t)$ for $T=0.01$. 
The dashed line indicate $\Delta_x/2T$, with $T=0.01$ the thermal 
bath temperature.}
\end{figure}\noindent
In Fig.~2 we analyze the FDR for the longitudinal direction: 
Fig.~2(a) shows 
the  quadratic mean displacement, $\Delta_y$, and  
Fig.~2(b) shows the response, $\chi_y$. In Fig.~2(c) 
we obtain the corresponding FDR. 
We observe that the equilibrium FDT applies for $\Delta_x(t)< 0.05 r_p^2$ 
at the bath temperature $T$. The plot does not have a constant slope
for larger displacements. Similar 
results are obtained in the plastic flow regime. 
\begin{figure}
\centerline{
\includegraphics[width=8.5cm]{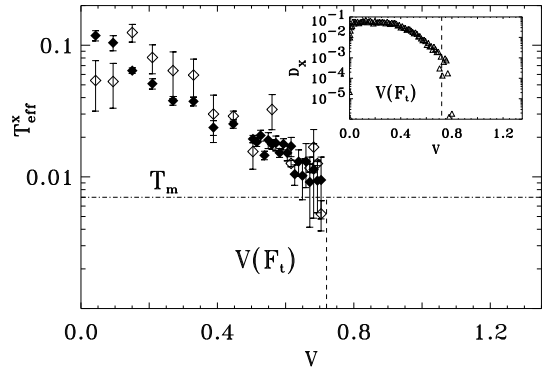}}
\caption{Transverse effective temperature $T_{\tt eff}^x$ vs voltage $V$ 
for $A_p=0.35$, $T=0$ and $n_v=0.07$, using the diffusion relation 
(filled diamonds) and a generalized Kubo formula (open diamonds). Inset: 
Transverse diffusion constant $D_x$ vs $V$. Dashed lines indicate the 
transverse freezing transition at $F=F_t$ and the dash-dotted line 
indicate the melting temperature 
of the unpinned system $T_m \approx 0.007$.}
\end{figure}\noindent

In Fig.~3 we show the calculated transverse effective temperature, 
$T_{\tt eff}^x$, for $T=0$
as a function of voltage ({\it i.e.}, average velocity, $V$). 
We observe that above the critical force, $T_{\tt eff}^x$, is a 
decreasing function of $V$ that reaches a value close to 
the equilibrium melting 
temperature of the unpinned system, $T_m \approx 0.007$ \cite{tmelting}, when 
the system approaches the transverse freezing transition at $F=F_t$
(obtained from the vanishing of the transverse diffusion 
$D_x$, shown in the inset). 
It becomes very difficult to compute numerically $T_{\tt eff}^x$ 
for driving forces $F>F_t$, 
since $\Delta_x$ and $\chi_x$ are bounded at $T=0$ 
while for finite $T$ there are very long relaxation times involved.
Therefore, we leave the interesting case of obtaining $T_{\tt eff}(T)$  
for $F>F_t$ for future study. The opposite limit $F\to 0$ is also 
interesting. We shall report on FDT measurements when $F=0$, 
in the Bragg and vortex glasses, elsewhere.

\begin{figure}
\centerline{
\includegraphics[width=8.5cm]{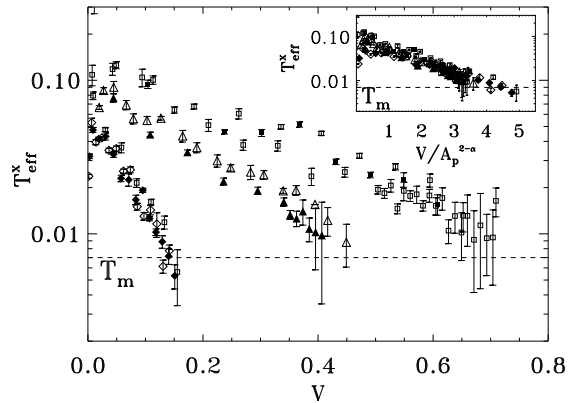}}
\caption{Transverse effective temperature $T_{\tt eff}^x$ vs voltage $V$, for 
different values of pinning amplitude $A_p$, and vortex density $n_v$, 
at $T=0$. $A_p=0.35$, $n_v=0.05$ ($\blacksquare$), $A_p=0.35$, $n_v=0.07$ 
($\square$), $A_p=0.2$, $n_v=0.05$ ($\blacktriangle$), $A_p=0.2$, $n_v=0.07$ 
($\vartriangle$), $A_p=0.1$, $n_v=0.05$ ($\lozenge$), $A_p=0.1$, $n_v=0.07$ 
($\blacklozenge$). The inset shows $T_{\tt eff}^x$ vs $V/A_p^{2-\alpha}$, with 
$\alpha=0.5$ }
\end{figure}\noindent

In Fig.~4 we show the dependence of 
$T_{\tt eff}^x$ with 
pinning amplitude, $A_p$, and vortex density, $n_v$. 
In all the cases we 
observe that $T_{\tt eff}^x \rightarrow T_m$ when $F \rightarrow F_t$; 
even when $F_t$ depends on $A_p, n_v$. 
The inset of Fig.~4 shows that we can 
approximately collapse all the curves plotting 
$T_{\tt eff}^x$ vs $V/A_p^{2-\alpha}$ with 
$\alpha \sim 0.5$. The shaking temperature for a 
single vortex is expected to satisfy this scaling with $\alpha=0$~\cite{SV}. 
The same result should 
apply in the limit of non-interacting vortices or incoherent motion \cite{KV}. 
In the opposite case of motion of a rigid lattice
we can apply the one particle result to a Larkin-Ovchinikov 
correlation volume where the pinning force summation gives an 
effective pinning amplitude $\sqrt{A_p}$, and therefore 
$\alpha=1$ in this limit. 
It is noteworthy that the value we find is intermediate between 
these two limits. 

We now show that the same $T^x_{\tt eff}$ can be obtained 
from a different observable. 
In particular from experimentally accessible quantities like the 
transverse resistivity and the voltage fluctuations. 
If $T^x_{\tt eff}$ is well defined in a given time scale 
we can expect a generalized Kubo formula to hold~\cite{Yoshino}, 
\begin{equation}
R_x(t)=\frac{1}{k_B T^x_{\tt eff}(t)} 
\int^t_0 dt' \langle V_x(t) V_x(t')\rangle \; ,
\end{equation}
where $R_x(t) =\langle \frac{dV_x(t)}{d\epsilon} \rangle_{\epsilon=0}
-\langle \frac{dV_x(0)}{d\epsilon} \rangle_{\epsilon=0} $ 
is the linear transverse resistance. If we now make a parametric
plot of $R_x(t)$ vs. $\int^t_0 dt' \langle V_x(t) V_x(t')\rangle$
we again find a linear slope equal to $1/T$ for $t\ll   r_p/v$ and 
a second linear slope of $1/T_{\tt eff}^x$ for $t > r_p/v$.
In Fig.~2(a) we compare 
the two effective transverse temperatures obtained 
using Eqs.~(4) and (5).
We see that they are similar within the error bars.
The shaking temperature, $T_{\tt sh}$, defined in Eq.~(3) of Ref.~\cite{KV} 
is proportional to the time integral of the correlation function of the 
pinning force ${\bf F}_{\tt pin}=\sum_{i,p}{\bf f}_{ip}(t)$. $T_{\tt sh}$
can be obtained from Eq.~(6) if we replace  $R_x(t)$ with the single
vortex value $R_0=1/\eta$ and the integral of the $V_x(t)$ correlation
function is taken for all $t$ [since for the transverse direction 
$V_x(t)\propto F_{\tt pin}^x(t)$]. 
In other words,  $T_{\tt sh}$ of \cite{KV}
corresponds to taking the average slope in the parametric plot of  
the generalized Kubo formula [or in the parametric plot of the FDR shown
in Fig.~1(c)], see also~\cite{shear_foam}. 
The approach followed here permits defining an effective
temperature which takes into account all the information on its time scale
dependence that allows for a thermodynamic interpretation of $T_{\tt
eff}$. 
In this way we see clearly that there is a non-trivial
value of the transverse effective temperature
$T^x_{\tt eff}$  for time scales $t > r_p/v$. Since one drives the 
system in the longitudinal direction $y$, the condition of slow
motion needed to prove the thermodynamic properties in $T_{\tt eff}^y$
is not matched. Thus, it is no surprise that we do not 
find the same value of $T_{\tt eff}$ in the longitudinal 
direction~\cite{note,teff_gran2}.
Furthermore, we find that the short-range correlations of the moving
fluid are important and give a non-trivial dependence with disorder
strength $A_p$.
Besides this, we have demonstrated that in the moving fluid phase
$T^x_{\tt eff}$ satisfies two important results of~\cite{KV}:
(i) additivity of temperatures, $T^x_{\tt eff}(T)=T^x_{\tt eff}(0)+T$ and
(ii) dynamic freezing occurs when $T^x_{\tt eff}(T)=T_m$.
From the analysis of Eq.~(6), we  note
that $\langle V_x(t)V_x(0)\rangle$ can be obtained from transverse
voltage noise measurements (from $|V_x(\omega)|^2$) 
and $R_x(t)$ from time dependent measurements of the transverse
resistivity. It  will therefore be interesting to have
experimental measurements of $T_{\tt eff}$ to test quantitatively the
dynamic freezing transition.  Finally, we stress that a complete dynamic
theory of the moving vortex system has to capture the features  
here described.

We acknowledge discussions with J. Kurchan, T. Giamarchi, P. Le Doussal,
M. C. Marchetti and V. M. Vinokur.
We acknowledge support from the Argentina-Francia international 
cooperation SETCIP-ECOS, project A01E01,
from  ANPCyT (PICT99-03-06343) and Conicet, and from the Director,
Office of Adv.\ Scientific Comp.\ Res., Div.\ of Math., Information, and 
Comp.\ Sciences, U.S.\ DoE contract DE-AC03-76SF00098.


\begin{thebibliography}{}
\bibitem{note} We note that also in sheared fluids an effective temperature 
has been obtained only 
in the direction perpendicular to the driving force~\cite{shear}.
\end{thebibliography}
\end{document}